\documentclass[journal, a4paper]{IEEEtran}

\usepackage{url}
\usepackage{graphicx}
\usepackage{stfloats}
\usepackage{amsmath}
\usepackage{amssymb}
\usepackage{sidecap}
\usepackage{mathtools}
\usepackage{makecell}

\newcommand{\bfrac}[2]{\displaystyle \frac{#1}{#2}}

\begin{document}

\title{Representation of arbitrary rational functions and polynomials by non-Foster Resistor-Capacitor-Inductor
networks}

\author{A. J. Mackay
\thanks{A. J. Mackay is principal consultant at Universal Electromagnetics Ltd., UK. (e-mail: andrew@ueltd.co.uk).}
\thanks{The author acknowledges funding via QinetiQ from DSTL.}}
\markboth{ }{ }

\maketitle

\begin{abstract}
The Bott-Duffin theorem
 shows that any arbitrary positive real rational function of finite order,
 representing a passive network, 
can be represented by a finite number of positive value resistors, capacitors and inductors. It is shown
here that
if the
positivity restriction is lifted, any rational function or polynomial of any order may be represented by
a network of (possibly negative) inductors, capacitors and resistors. When there are no repeated roots this
is a sum of simple sub-circuits each of which contains no more than four components. When there are
repeated roots a more complex circuit is required, here featuring iterated Wheatstone bridges of special form.
\end{abstract}


\IEEEpeerreviewmaketitle

\section{Introduction}

The use of equivalent circuit representations is fundamental to much of electrical engineering and the
description of metamaterials
for use at radio, microwave and optical frequencies. The celebrated
Bott-Duffin theorem \cite{Youla}
shows that any positive real rational function is
synthesisable as
a network containing only positive inductors (L), capacitors (C) and resistors (R).
Such rational functions have special properties; for example any poles and zeros in the complex $s-$ plane 
can lie only in the left half of the complex plane and the orders of the numerator and denominator polynomials representing
the rational function can differ by at most one.

In recent times there
has been significant interest in the representation of active networks, where it is possible to represent impedance
functions by non-Foster networks of inductors (L), capacitors (C) and resistors (R) which may be non-positive. Such
networks may be unstable and/or demonstrate gain.
Here it is shown that any polynomial or rational function of finite order on the field of real
numbers
can be represented by a finite network of (possibly negative) LCR values. Negative values may be synthesised
at radio and microwave frequencies using discrete amplifiers and negative impedance converters. At mm-wave and
optical frequencies negative values may be generated through parametric amplification and time-modulation of the
relative permittivity and permeability.

When there are no 
repeated roots the LCR values are determined explicitly and the
the total component count scales with the order of the rational
function or polynomial. General rational functions and polynomials with repeated roots may be expressed using
iterated Wheatstone bridge sub-circuits to model the degenerate terms. 

Although these results  are formal with no consideration of stability it is possible that the non-Foster representation
may have value, both for functional analysis and as a tool for the circuit analysis of active structures in the optical
and microwave domains.

\section{Assertion} 

Suppose a function $f(s)$, real for $s$ real, can be represented by the rational form,
\begin{equation}
f(s) = \frac{a_0s^N+a_1 s^{N-1} + \, . \, . \, . a_N}
            {b_0s^M +b_1 s^{M-1} + \, . \, . \, . b_M}
\label{ratform}
\end{equation}
for real valued coefficients $a_i$ and $b_i$. Let the orders $N\ge 0$ and $M\ge 0$ of the polynomials in the
numerator and the denominator be arbitrary.
When $s=j\omega$ for a real frequency $\omega$, $f(s)$ can represent the point-to-point impedance of a network 
comprising a finite number of inductors, capacitors and resistors, $(L_i, C_j, R_k)$ provided these may take arbitrary sign.

The assertion is a generalisation of the Bott-Duffin theorem for positive rational functions expressible as a finite network
of positive $L,C,R$ values. A necessary requirement for positive rational functions is that $|N-M| \le 1$
 and the number of required $LCR$ values is
 bounded by an exponential function of $N$ for large $N$. In the generalisation for non-Foster
networks the circuit representation shows that the number
of required $LCR$ values is bounded by a linear function of $N$ and $M$ when there are no repeated roots
and a log-linear function of $N$ and $M$ when there are repeated roots of arbitrary degeneracy.

\section{Proof}

\begin{figure}[htp]
\centerline{\includegraphics[width=0.10\textwidth]{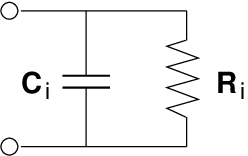}}
\caption{A single RC unit from the first sum $T^{(1)}$.} 
\label{RC}
\end{figure}

\begin{figure}[htp]
\centerline{\includegraphics[width=0.40\textwidth]{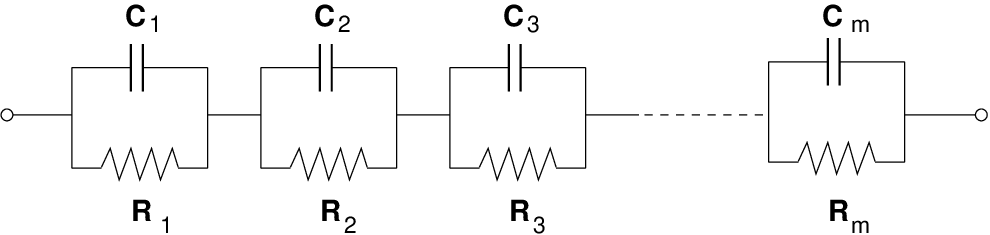}}
\caption{Multiple RCS units representing the first sum $T^{(1)}$} 
\label{RCchain}
\end{figure}

\begin{figure}[htp]
\centerline{\includegraphics[width=0.40\textwidth]{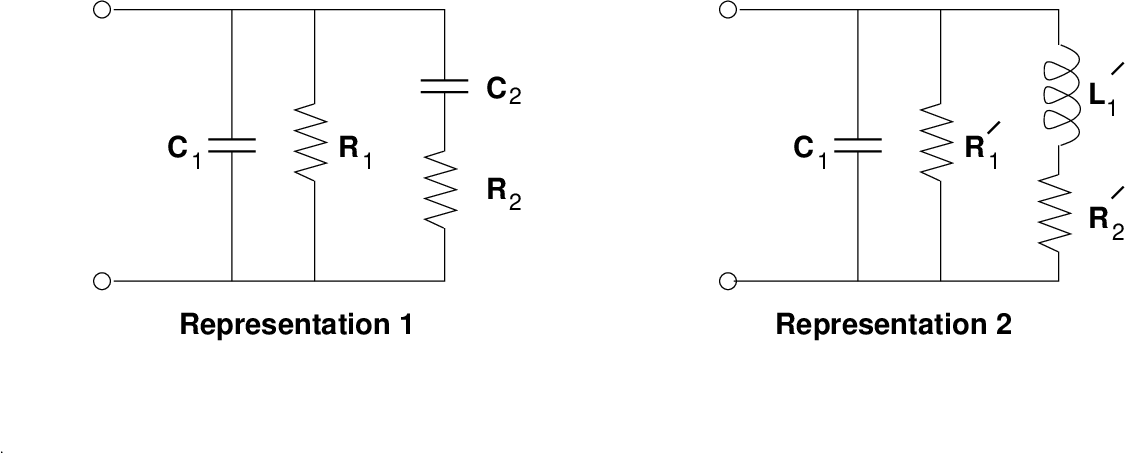}}
\caption{A single equivalent circuit unit from the second sum, $T^{(2)}$} 
\label{RRC_type2}
\end{figure}

On the field of real numbers, $f(s)$ may be expressed according to the fundamental theorem of algebra by,
\begin{equation}
f(s)= P(s) + \sum_{i=1}^m \sum_{r=1}^{j_i}  \frac{D_{ir}}{(s-d_i)^r}
           + \sum_{i=1}^n \sum_{r=1}^{k_i}  \frac{A_i s + B_i}{(s^2+a_i s+ b_i)^r}
\label{fundrep}
\end{equation}
where $P(s)$ is a polynomial, zero if $N<M$, and all coefficients are real-valued. The roots of the
quadratic terms are assumed complex and $j_1+j_2+\, ... \,+j_m+ 2(k_1+k_2+\, ... \,+k_n)=M$.
 The first part of
the proof is restricted
 to the case where there are no repeated roots, i.e. where
$k_i=j_i=1$ and the polynomial $P(s)$ has no repeated roots. So,
\begin{equation}
f(s)= P(s) + \sum_{i=1}^m  \frac{D_i}{s-d_i}
           + \sum_{i=1}^n  \frac{A_i s + B_i}{s^2+a_i s+ b_i}
\end{equation}
with all coefficients real-valued and $P(s)$ real when $s$ is real. First consider a single term of the first sum,
\begin{equation}
T_i^{(1)}=\frac{D_i}{s-d_i}
\end{equation}
When $T_i$ represents the impedance of a circuit, the circuit representation is shown in figure \ref{RC},
where the capacitor $C_i$ and resistor $R_i$ take the values,
$C_i=1/D_i$ and $R_i=-D_i/d_i$.
Consequently, the sum,
\begin{equation}
T^{(1)}=\sum_{i=1}^m  T_i^{(1)}
\end{equation}
is represented by the equivalent circuit in figure \ref{RCchain},
Clearly there are $2m$ components where $m \le M$.
Now consider a term from the second sum,
\begin{equation}
T_i^{(2)}=\frac{A_i s + B_i}{s^2+a_i s+ b_i}
\label{Ti2}
\end{equation}
such that
\begin{equation}
T^{(2)}=\sum_{i=1}^n T_i^{(2)}
\end{equation}
Each term has two possible equivalent circuit representations shown in figure \ref{RRC_type2}.
where, for configuration 1,
\begin{eqnarray}
C_1 &=& 1/A_i    \nonumber \\
R_1 &=& B_i/b_i  \nonumber \\
C_2 &=& \bfrac{B_i(A_i a_i-B_i)-b_i A_i^2}{A_i B_i^2} \nonumber \\
R_2 &=& \bfrac{B_i A_i^2}{B_i(A_i a_i-B_i) - b_i A_i^2}
\end{eqnarray}
and for configuration 2,
\begin{eqnarray}
C_1 &=& 1/A_i    \nonumber \\
R_1^\prime &=& \bfrac{A_i^2}{A_i a_i -B_i}  \nonumber \\
L_1^\prime &=& \bfrac{A_i^3}{A_i^2 b_i - B_i(A_i a_i-B_i)} \nonumber \\
R_2^\prime &=& -\bfrac{B_i A_i^2}{B_i(A_i a_i-B_i) - b_i A_i^2}
\end{eqnarray}
These equivalent circuit units are also chained together to represent $T^{(2)}$. There are $4n$ components in this chain,
where $n \le M/2$.

The remaining part of the expression is the polynomial $P(s)$ of maximum order $N-M$. In order to find an equivalent circuit
for this consider the reciprocal function, $g(s) = 1/P(s)$. This is clearly another rational function that takes the 
form,
\begin{equation}
g(s)=  \sum_{i=1}^{m^\prime}  \frac{D^\prime_i}{s-d_i^\prime}
       + \sum_{i=1}^{n^\prime}  \frac{A_i^\prime s + B_i^\prime}{s^2+a_i^\prime s+ b_i^\prime}
\end{equation}
with no polynomial term. An equivalent circuit has just been found for this, with primed real-valued constants replacing un-primed ones.

\begin{figure}[htp]
\centerline{\includegraphics[width=0.40\textwidth]{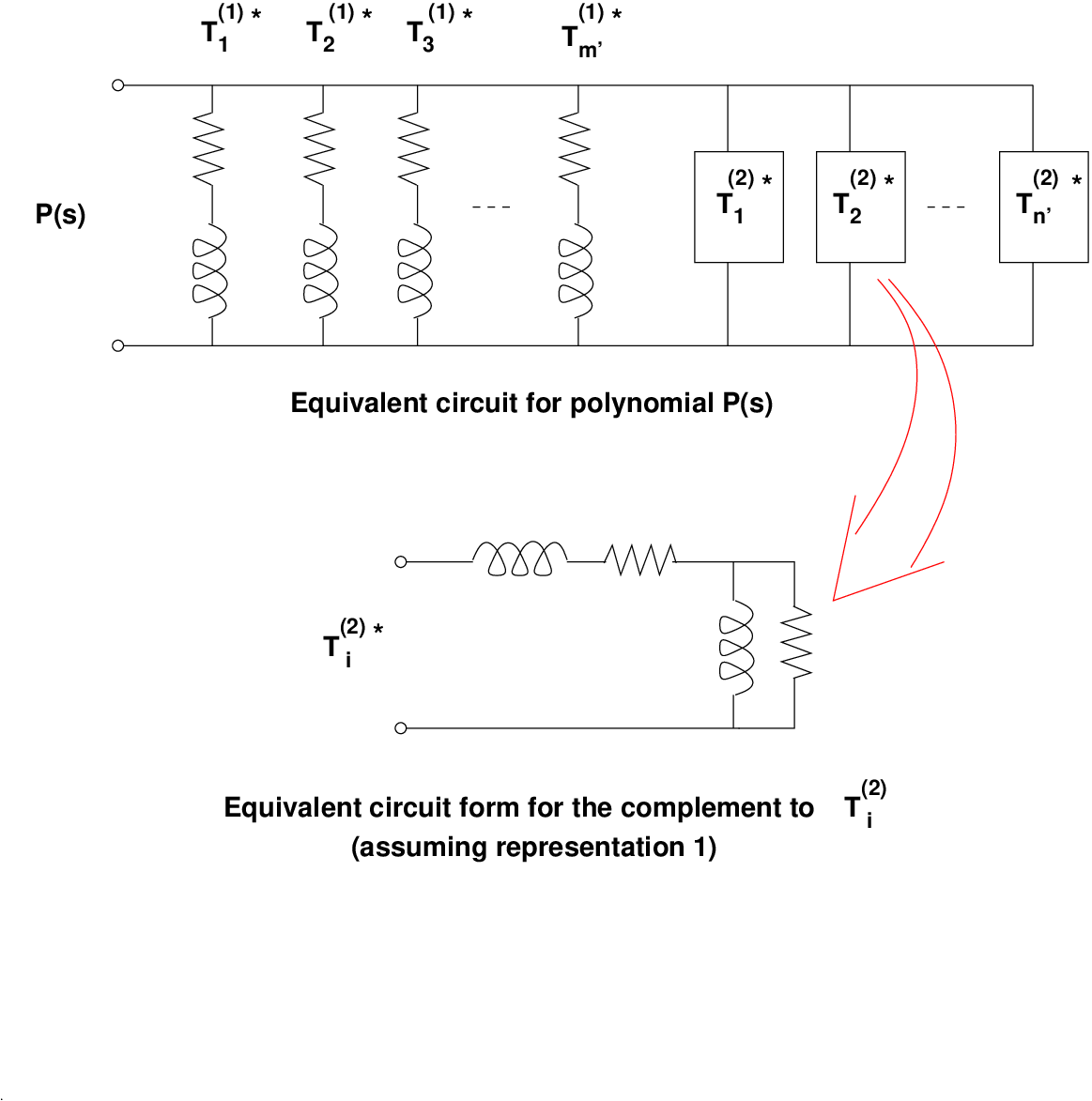}}
\vspace*{-0.2in}
\caption{Equivalent circuit representing the polynomial $P(s)$} 
\label{P_form}
\end{figure}

Finally, since $g(s)$ represents an equivalent circuit of a ladder network,
$1/g(s)$ has a realisable complementary form obtained by the impedance-to-admittance rule, where capacitors are replaced by
inductors (and vice-versa) and series arrangements of components are replaced by parallel ones (and vice-versa). Consequently,
$P(s)$ can be realised by the component arrangement illustrated in figure \ref{P_form}. In the figure, the superfix $*$ 
signifies the equivalent circuit of elements of the complementary network. 

The equivalent circuit representation for $f(s)$ is the series sum for $P(s)+T^{(1)}(s)+T^{(2)}(s)$. This completes the proof when there is no degeneracy.

\section{A simple non-degenerate example}

\begin{figure}[h]
\centerline{\includegraphics[width=0.15\textwidth]{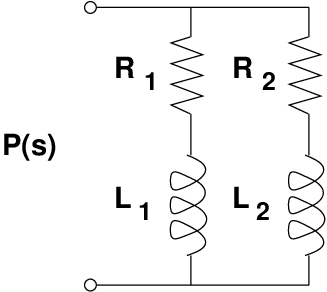}}
\caption{Equivalent circuit for a quadratic with distinct real roots} 
\label{P_quad}
\end{figure}

Consider the quadratic polynomial $P(s)=(s+a)(s+b)$ with two real roots for which $a \ne b$.
 This cannot be a positive real function since $|N-M|>1$
and has the equivalent
circuit given by the parallel combination of two RL series components \eqref{P_quad},
Using the previously established rules, $P(s)$ represents the impedance function where,
\begin{eqnarray}
L_1 &=& b-a \nonumber \\
L_2 &=& a-b \nonumber \\
R_1 &=& a(b-a) \nonumber \\
R_2 &=& b(a-b)
\end{eqnarray}
Observe that in the limit as $a\rightarrow b$, all component values tend to zero establishing that when
the function has repeated roots this equivalent circuit is invalid.

\section{Polynomials and rational functions with repeated roots}

\begin{figure}[h]
\centerline{\includegraphics[width=0.45\textwidth]{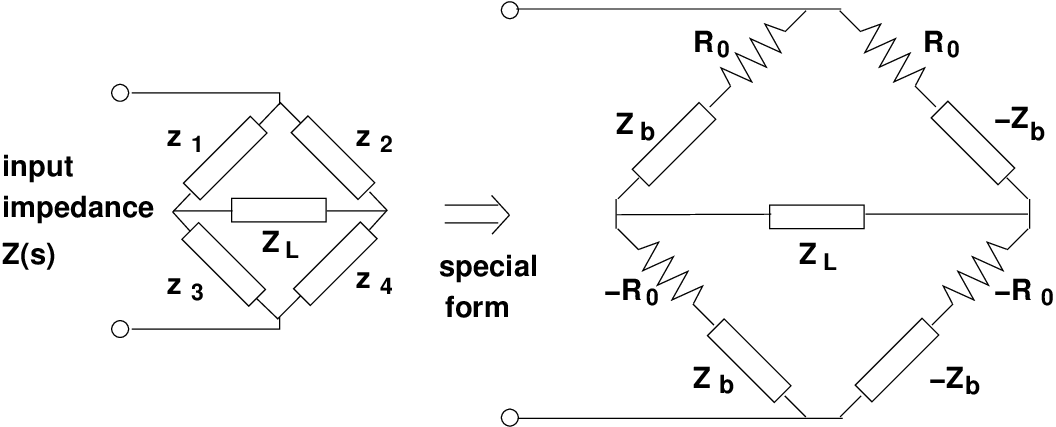}}
\caption{Equivalent circuit synthesis of repeated roots using a special Wheatstone bridge} 
\label{Wheat}
\end{figure}

When there are repeated roots this network construction fails though finite equivalent circuits still
exist. 
The repeated root terms in the expression for $f(s)$ can be determined
using  a special unbalanced Wheatstone bridge as shown in figure \ref{Wheat}.
The input impedance $Z(s)$ of a general bridge is given by,
\begin{equation*}
\hspace*{-0.15in}
Z(s) = \frac{z_1z_3(z_2+z_4)+z_2z_4(z_1+z_3)+Z_L(z_1+z_3)(z_2+z_4)}
            {(z_1+z_2)(z_3+z_4) + Z_L(z_1+z_2+z_3+z_4)}
\end{equation*}
which may be most easily derived using the Extra Element Theorem \cite{Middlebrook} as shown in \cite{Mehmet}.
Using the special form on the right of figure \ref{Wheat} the first two terms in the numerator and second term
in the denominator vanish and the expression simplifies to,
\begin{equation}
Z(s)=\frac{(Z_b(s))^2}{(R_0)^2} \, Z_L(s)
\label{wform}
\end{equation}
which may be used to determine equivalent circuits both of polynomial terms and of partial fractions with
repeated roots.\footnote{This special form is not new but the author was unable to find a definitive reference
 in online
searches.} There are many ways this can be done and the representations below are not unique.
Provided that at any point $Z_b(s)$ and $Z_L(s)$ are represented by realisable equivalent circuits (either 
degenerate or non-degenerate), $Z(s)$ represents a new equivalent circuit which may in turn be taken as
a new input to $Z_b(s)$ or $Z_L(s)$ in a subsequent generation. Equation \eqref{wform} is applied to \eqref{ratform}
but rather than using the form \eqref{fundrep} employ the representation,
\begin{equation*}
\begin{aligned}
 f(s)= &  \frac{(s-\alpha_1)^{p_1}(s-\alpha_2)^{p_2}...(s-\alpha_u)^{p_u}}
           {(s-\beta_1)^{q_1}(s-\beta_2)^{q_2}...(s-\beta_v)^{q_v}} \\
 \times &  \frac{(s^2+\gamma_1 s+ \epsilon_1)^{p^\prime_1} (s^2+\gamma_2 s+ \epsilon_2)^{p^\prime_2}...
            (s^2+\gamma_w s+ \epsilon_w)^{p^\prime_w}}
           {(s^2+\xi_1 s+ \eta_1)^{q^\prime_1}(s^2+\xi_2 s+ \eta_2)^{q^\prime_2}...
            (s^2+\xi_x s +\eta_x)^{q^\prime_x}}
\end{aligned}
\end{equation*}
where all coefficients are real, the quadratic terms have complex roots and
\begin{eqnarray}
p_1+p_2+ \, ... \, +p_u + 2(p^\prime_1+p^\prime_2+ \, ... \,+p^\prime_w) = N \nonumber \\
q_1+q_2+ \, ... \, +q_v + 2(q^\prime_1+q^\prime_2+ \, ... \, +q^\prime_x) = M \nonumber
\end{eqnarray}
with repeated real roots of order $p_1$, $p_2$ ... $p_u$ in the numerator,
repeated real roots of order $q_1$, $q_2$ ... $q_v$ in the denominator,
repeated complex conjugate roots of order $p^\prime_1$, $p^\prime_2$ ... $p^\prime_w$ in the numerator
and repeated complex conjugate roots of order $q^\prime_1$, $q^\prime_2$ ... $q^\prime_x$ in the denominator.

The function $f(s)$ is now partitioned as $f(s)=g(s)h(s)$  where $h(s)$ has repeated roots in the numerator and denominator
which are all even and $g(s)$ has no repeated roots. For each factor, if $p_i$, $q_i$, $p^\prime_i$ or $q^\prime_i$
is even then this factor is retained in $h(s)$. If, for any factor, $p_i$, $q_i$, $p^\prime_i$ or $q^\prime_i$
is odd then $p_i$, $q_i$, $p^\prime_i$ or $q^\prime_i$ may be increased  or decreased by one
in the expression for $h(s)$ provided this additional factor is included
in the numerator or denominator of $g(s)$ maintaining the equality $f(s)=g(s)h(s)$. In so doing,
$g(s)$ remains non-degenerate with an equivalent circuit given previously and
it is valid to take $Z_L(s)=g(s)$ and $(Z_b(s)/R_0)^2=h(s)$. Since all the factors in
$h(s)$ are even, $Z_b(s)$ is a rational function whose factors may in turn have even or odd order and the 
algorithm may be applied repeatedly until all degeneracy is removed.

\section{Some simple degenerate examples}

First consider 
the superinductance function $Z(s) =K s^2$, for constant $K>0$
 with a double zero at the origin. This
may be synthesised with
$Z_b(s) =sL$ representing 
an inductor and $Z_L(s)=R_L$ representing a resistor,
$R_0$, $R_L$ and $L$ assumed positive. This result is well established in the literature with
 $Z(s)=(L/R_0)^2 R_L s^2$.
The supercapacitance function $Z(s) = K/s^2$ can be constructed in like manner where $Z_b(s)=1/(sC)$.
Higher order terms with $Z(s) = K s^p$, $p>2$, may be constructed by replacing $Z_b(s)$ and/or $Z_L(s)$
 by $Z(s)$ determined with a smaller value of $p$, from a ``previous
generation''.

Polynomial terms of the form $Z(s)= K(s+a)^p$ for $p>1$
and arbitrary real constant $a$ may be generated either by expansion and summation of terms
of the form $K_i s^i$ ($0 \le i\le p$, $K_i$ real constants), or by employing $Z_b(s)=R_b+sL$
 representing a resistor $R_b$
 (possibly negative)
 in series with an inductor $L$   followed (for $p>2$) by higher generation replacement of
$Z_b(s)$ and/or $Z_L(s)$ by $Z(s)$ from a previous generation. Similarly, elements
of the form $Z(s)= (T_i^{(1)})^p \equiv K(s+a)^{-p}$ for $p>1$
 can be constructed using the same bridge with $Z_b(s)=1/(1/R_b +sC)$
representing a resistor and capacitor in parallel as illustrated in figure \ref{RC}.

Terms like $(As+B)/(s^2+as+b)^p$ can be constructed by factorising as $g(s)h(s)$  in two possible ways.
When $p$ is odd, write
$g(s)=(As+B)/(s^2+as+b)$ and $h(s)= 1/(s^2+as+b)^{p-1}$  
When $p$ is even, write $g(s)=As+b$  and $h(s)= 1/(s^2+as+b)^p$. As before this allows the identification,
 $f(s)=Z_L(s)$ and $g(s)=(Z_b(s)/R_0)^2$.
 The order of the degeneracy is reduced by a factor of two
and the degeneracy reduction can be continued iteratively using the same method.
Note that it is also possible to decompose $(As+B)/(s^2+as+b)^p$ as a sum of the terms 
$As/(s^2+as+b)^p$ and $B/(s^2+as+b)^p$ and employing the algorithm on each
of the two terms separately if this is required.

It is clear that any polynomial or rational function with degenerate roots can be expanded using iterated bridges
in many different ways, though in all cases the component count is much larger
 than for non-degenerate functions with no repeated roots.
 For a general rational function, degeneracy
can approach the order of the rational function $p-1 = O(\max(N,M))$.
The proposed Wheatstone iteration employs the quadratic term
$(Z_b(s))^2$ so that at each iteration the degeneracy halves.
 The degeneracy is an exponential function of the iteration
number so that the component count is bounded by a factor that scales as
$\max(N,M) \log (\max(N,M))$. This is made clear in the examples below where,
if $Z_b$ is employed every generation, the $\max(N,M) \log (\max(N,M))$ bound is
valid.

\section{Two more complicated degenerate examples and the component count bound}

Here consider the rational function,
\begin{equation}
f(s)\equiv Z(s) =\frac{(a_1 s +b_1)^3 (a_2s+b_2)}{(c_1 s + d_1)^2}
\end{equation}
Choose $Z_b(s)=R_0 (a_1 s +b_1)/(c_1 s + d_1)$ and $Z_L(s)=(a_1 s+b_1)(a_2 s+b_2)$. The $Z_L(s)$ is non-degenerate
and its equivalent circuit was represented in the first example. The term for $Z_b(s)$ is also non-degenerate with a
prior representation. In fact, $Z_b(s)=R_1+R_2 ||(1/sC)$, i.e. a resistor $R_1$ in series with the parallel combination
of another resistor $R_2$ and capacitor $C$. It is straightforward to determine the values of these components
given arbitrary (real) coefficients in $f(s)$.

Next consider the rational function,
\begin{equation}
f(s)\equiv Z(s) =\frac{(a_1 s^2 +b_1 s +c_1)^3 (a_2s+b_2)}{(a_1^\prime s^2 + b_1^\prime s +c_1^\prime)^4}
\end{equation}
where neither of the quadratics in the numerator and denominator have real roots.
Choose $Z_{b,2}(s)=R_0 (a_1 s^2 +b_1 s +c_1)^2/(a_1^\prime s^2 + b_1^\prime s +c_1^\prime)^2$ and
$Z_{L,2}=(a_2 s+b_2)/(a_1^\prime s^2 + b_1^\prime s +c_1^\prime)$ where the subscript `2' refers to a
bridge at generation 2. The $Z_{L,2}(s)$ is non-degenerate and takes one
of the two representations for $T_i^{(2)}$ in \eqref{Ti2}. The $Z_{b,2}(s)$ term is simply the bridge impedance
$Z(s)$ associated with the first generation
 $Z_{b,1}(s)=R_0(a_1 s^2 +b_1 s +c_1)/(a_1^\prime s^2 + b_1^\prime s +c_1^\prime)$
(a non-degenerate bi-quadratic) and $Z_{L,1}=R_0$.

In general, whenever there is a degeneracy $p$ for large $p$ the most efficient decomposition is
to halve the degeneracy using the $Z_b$ term.
When $p$ is odd, subtract or add one by use of the $Z_L$ term and then halve the remainder using the $Z_b$ term.
This system may be applied with degeneracy in either the numerator or denominator of a rational function and where
the factor is either quadratic or linear in $s$. The order reduction in the iterative process is
shown symbolically for three examples of large $p$, without reference to the particular factors in 
the numerator and denominator, in the diagram below.
\begin{equation*}
\begin{tabular}{llllllllll}
p=15=16-1                        & & p=17=16+1                        & & p=103=102+  1                      & & \\
   \hspace*{0.4in} $\downarrow$  & &  \hspace*{0.4in} $\downarrow$    & &     \hspace*{0.4in} $\downarrow$   & & \\
   \hspace*{0.4in} 8             & &  \hspace*{0.4in} 8               & &     \hspace*{0.4in}  51=52-1       & & \\
   \hspace*{0.4in} $\downarrow$  & &  \hspace*{0.4in} $\downarrow$    & &     \hspace*{0.65in} $\downarrow$  & & \\
   \hspace*{0.4in} 4             & &  \hspace*{0.4in} 4               & &     \hspace*{0.65in} 26            & & \\
   \hspace*{0.4in} $\downarrow$  & &  \hspace*{0.4in} $\downarrow$    & &     \hspace*{0.65in} $\downarrow$  & & \\
   \hspace*{0.4in} 2             & &  \hspace*{0.4in} 2               & &     \hspace*{0.65in} 13=12+1       & & \\
   \hspace*{0.4in} $\downarrow$  & &  \hspace*{0.4in} $\downarrow$    & &     \hspace*{0.90in} $\downarrow$  & & \\
   \hspace*{0.4in} 1             & &  \hspace*{0.4in} 1               & &     \hspace*{0.90in} 6             & & \\
                                 & &                                  & &     \hspace*{0.90in} $\downarrow$  & & \\
                                 & &                                  & &     \hspace*{0.90in} 3=2+1         & & \\
                                 & &                                  & &     \hspace*{1.05in} $\downarrow$  & & \\
                                 & &                                  & &     \hspace*{1.05in} 1             & & \\
\end{tabular}
\end{equation*}
Here, the arrows represent a halving of the degeneracy using $Z_b$. Whenever there is
a `+' or a `-' a term, the degeneracy is converted from an odd to an even number with the conversion factor
put into $Z_L$. The number of times a factor is required in $Z_L$
is bounded by the number of times the order can be halved using $Z_b$. This shows that a bound on
the component count
scales as $\max(N,M) \log (\max(N,M))$.

\section{A few notes on stability}

While the results in this paper are entirely formal, it is worth making a few fundamental remarks concerning stability.
Firstly, if the function $f(s)$ represents an active impedance then it 
has either poles or zeros (or both) in the right hand half of the complex $s-$plane.
 If it has such poles then it is open-circuit unstable and if it has such zeros then it is short-circuit unstable.
Stability is defined only with reference to a terminating load. However, if $f(s)$ is stable under load then
every sub-circuit of its equivalent circuit should also be stable and this depends on the non-unique circuit
representation.
 Moreover, any negative LCR value generated
by a negative impedance converter must be synthesised using amplifiers and sub-circuits which should themselves be
 internally
stable under the in-circuit load conditions. Such devices must either be open circuit unstable or short circuit
unstable, designed for the load environment in which they are placed.

Formally there is no requirement for stability and indeed an impedance function represented by
a general polynomial or rational function
may be unstable under a given load irrespective of its circuit representation. However,
 there will be applications where stability
is required. If so then to summarise: (1)  the function $f(s)$ must be stable under load, (2)
every sub-circuit of its equivalent circuit representation must be stable under that load and
(3) the negative impedance converters used to generate the negative LCR values must also be stable under
the operating conditions. The first and third requirements can be taken
as givens, but the second requirement depends on the circuit representation.
An analysis of the circuit topologies constructed in this article and how they affect internal stability 
remains to be explored.


\begin{thebibliography}{1}
\bibitem{Youla} D. C. Youla, {\it {Theory and synthesis of linear passive time-invariant networks}}. Cambridge
University Press , 2015.
\bibitem{Middlebrook} R. D. Middlebrook, {\it Null double injection and the extra element theorem} IEEE Trans Education
Vol. 32, No. 3,  pp 167-180, August 1989.
\bibitem{Mehmet} Mehmet Can,{\it Youtube video on the Extra Element Theorem}, \newline
 www.youtube.com/watch?v=mP2OWGAQDBI, August 2025.

\end{thebibliography}
\end{document}